\documentclass[sigplan,10pt,natbib,nonacm]{acmart}

\def\techreport{1}

\acmJournal{PACMPL}
\acmNumber{OOPSLA} 
\acmYear{2023}
\acmMonth{August}
\acmDOI{} 

\setcopyright{none}

\usepackage{amsmath}
\usepackage{amsthm}

\usepackage{url}
\usepackage{listings, lstautogobble}
\usepackage{xspace}

\usepackage{hyperref}

\if\techreport1
    \newcommand{\bosqueir}{\textsc{BosqueIR}\xspace}
    \newcommand{\bosque}{\textsc{Bosque}\xspace}
    \newcommand{\srclocation}{\url{https://github.com/BosqueLanguage/BosqueCore}}
\else
    \newcommand{\bosqueir}{\textsc{LangXIR}\xspace}
    \newcommand{\bosque}{\textsc{LangX}\xspace}
    \newcommand{\srclocation}{Omitted For Review}
\fi

\newcommand{\eg}{\hbox{\emph{e.g.}}\xspace}
\newcommand{\ie}{\hbox{\emph{i.e.}}\xspace}
\newcommand{\etc}{\hbox{\emph{etc.}}\xspace}
\newcommand{\vs}{\hbox{\emph{vs.}}\xspace}

\newcommand{\wrt}{\hbox{\emph{w.r.t.}}\xspace}

\newcommand{\cf}[1]{\texttt{#1}}

\newcommand\bnfalt{\;\;|\;\;}
\newcommand\bnfas{\;\;:=\;\;}

\usepackage{color}
\definecolor{purple}{RGB}{75, 0, 255}
\definecolor{cgreen}{rgb}{0.25,0.5,0.35} 

\lstdefinelanguage{bosque}{
keywords={concept, entity, datatype, typedecl, provides, field, switch, match, abstract, method, if, then, elif, else, function, return, true, false, none, let, var, in, requires, ensures, invariant, validate, recursive, using, of, this, $, pred, fn, ref, examples, for, defer, test, const, override, $return},
keywordstyle=\color{blue}\bfseries,
identifierstyle=\color{black},
alsoother={@},
sensitive=true,
comment=[l]{//},
morecomment=[s]{/*}{*/},
commentstyle=\color{cgreen}\bfseries,
stringstyle=\color{red}\ttfamily
}
 
\begin{document}


\title{Toward Programming Languages for Reasoning -- Humans, Symbolic Systems, and AI Agents}

\author{Mark Marron}
\affiliation{
  \institution{University of Kentucky}            
  \country{USA}                    
}
\email{mark.marron@protonmail.com}

\begin{abstract}
Integration, composition, mechanization, and AI assisted development are the driving themes in 
the future of software development. At their core these concepts 
are rooted in the increasingly important role of computing in our world, the desire to deliver functionality 
faster, with higher quality, and to empower more people to benefit from programmatic automation.
These themes, and how they impact the human developers driving them, are 
the foundations for the next generation of programming languages.
At first glance the needs of mechanization tools, AI agents, and human developers along with 
the various goals around development velocity, software quality, and software democratization 
are a broad and seemingly diverse set of needs. However, at their core is a single challenge that, 
once resolved, enables us to make radical progress in all of these areas. 

Our hypothesis is that, fundamentally, software development is a problem of reasoning about code 
and semantics. This is true for human developers implementing a feature, symbolic tools building 
models of application behaviour, and even for language based AI agents as they perform tasks. 
While the particular aspects of reasoning that each agent struggles with varies to some degree, 
they share many common themes and, surprisingly, most mainstream languages extensively employ 
(anti)features that make this task harder or infeasible! This paper proposes a novel approach to this 
challenge -- instead of new language features or logical constructs, that add more complexity to 
what is already a problem of complexity, we propose radical simplification in the 
form of the \bosque platform and language.
\end{abstract}



\maketitle

\section{Introduction}
\label{sec:intro}

The introduction and widespread use of structured programming~\cite{structuredprogramming} and abstract data 
types~\cite{adts} marked a major shift in how programs are developed. Fundamentally, the concepts and designs 
introduced by these programming methodologies simplified reasoning about program behavior by eliminating 
substantial sources of, usually entirely \emph{accidental}~\cite{silverbullet}, complexity. This allowed engineers 
to focus on the intent and core behavior of their code more directly and, as a result, produced a drastic 
improvements in software quality and ability to construct large software artifacts. Just as accidental complexity 
is an impediment to human understanding of a program it is also an impediment to applying formal reasoning techniques, 
and other mechanization, to software systems. Despite the inability of structured programming to fully bridge 
the chasm of formal mathematical analysis, issues with loop-invariants and mutation-frames among others prevented 
the practical use of deep verification techniques, it did provide the needed simplifications for reasoning about 
limited forms of program behavior and supported a golden age of IDE tooling and compiler 
development~\cite{muchnick,kennedyallen}. 

This paper takes the view that the next phase of software development and programming languages will be defined 
by (1) componentization, (2) AI assistance, and (3) mechanization. We are already well into the first of these 
transformations where, in the last decade, service based (cloud) applications~\cite{restphd,serverlessdef}, package ecosystems~\cite{semver,npm},  
and frameworks~\cite{react,express} have played critical roles in shaping how developers work. 
The second transformation, the rise of AI assisted development, arrived with the release of github's Copilot 
tool~\cite{copilot}. Suddenly, developers are looking at a landscape where they are curators of code generated by an AI agent, 
and integrating rich framework-functionality, in addition to writing application code directly!

In this new world, the ability for various agents and systems to understand and reason about application behavior
is critical. Humans must reason about increasingly complex integrated systems, AI agents need to ingest and operate 
reliably on complex tasks and codebases, and, with the increased demands of AI/human agents, symbolic tooling must 
be able to scalably process these systems. All three of these audiences have challenges when reasoning about code 
written in mainstream languages today. Some of the challenges are shared, mutable state is always hard, while some 
are specific to a reasoning agent, \eg symbolic analysis of iterative constructs or humans overlooking special case behaviors. 
However, all of the agents must share a common model for representing an application and, our hypothesis is that, foundationally 
the structure of this representation can be designed in a way to maximize the ability all three types of agents to 
reason about the semantics of an application.

This paper proposes a novel frame for conceptualizing the role and design of a programming language and software stack. 
Specifically, re-conceptualizing the programming language as the foundation for reasoning and mechanization, with development 
as a process of integration, validation, and composition. This task involves starting from first-principles and looking carefully 
at the impact each design choice has on the ability of agents to reason about and process code. Using this methodology we 
identify a set of seven major sources of complexity and ways, via language design, that they can be eliminated or minimized. 

Using this insight we develop an intermediate language and surface syntax, \bosqueir and \bosque respectively, which take the novel 
approach of language design by the removal of problematic feature and reduction in power of various constructs -- as opposed to the 
usual approach of introducing new language features or extending them with more powerful capabilities. This approach 
results in a symbolic reasoning friendly IR and a human/AI agent reasoning friendly surface syntax. Although seemingly 
simple this result has deep implications for the future of software-development, compilers, runtimes, and AI agents. 

To demonstrate some of the possibilities that a language, like \bosque, creates we look at a case studies of long standing and 
impactful problems, a decidable theory for small-model program validation, and the new and critical problem of effectively 
generating (correct) code that matches developer/user intents using AI agents. In 
both of these cases a highly effective solution can be constructed via the straight forward framing of the properties of interest 
and the ability to effectively reason about how a piece of \bosque code interacts with these properties.\\

\noindent
The contributions of this paper are:
\begin{itemize}
\item A novel approach for conceptualizing the role and design of a programming language and software stack that 
is focused on the language as a substrate for reasoning, tools, and agents.
\item An analysis and enumeration of language features that impede analysis 
and reasoning (\autoref{sec:complexity}).
\item An intermediate program representation, \bosqueir, designed explicitly to support mechanized 
reasoning and analysis (\autoref{sec:bsqir}) and a surface syntax, \bosque, optimized for humans and AI agents (\autoref{sec:bsqsrc}). 
\item Two case-study systems that demonstrate the viability and potential for this approach to language 
design in the future of software development (\autoref{sec:casestudies}).
\end{itemize}


\section{Complexity and Confusion}
\label{sec:complexity}

Our goal is to create a programming system that is optimized for reasoning -- by humans, symbolic analysis tooling, and AI agents 
(Large Language Model~\cite{gpt4} type agents in particular). The approach we take is to identify and remove features or concepts that complicate 
reasoning for our agents. Based on a range of experiences and sources including developer interviews, personal experience 
with analysis/runtime/compiler development, and empirical studies this section identifies seven major sources of complexity that can be addressed 
via thoughtful language design. These are sources of various bug categories, increase the effort required for a developer (or AI agent) 
to reason about and implement functionality in an application, and greatly complicate (or make it infeasible to) automatically reason about a program. 

\subsection{Reasoning Agents and Challenges}
\label{sec:reasoningagents}
We begin by looking at the three reasoning agent types we are considering -- human, symbolic, and Large Language Model (LLM) systems. There are 
common themes in what aspects of reasoning each of these agents find challenging but, given the 
differences in their inference modes, we also see how they have unique challenges as well.

\noindent
\paragraph{\bf Human Developers:} Humans are remarkably adaptable and are able to understand code using many modalities 
including symbolic modeling, dynamic tracing of execution on concrete values, inferences from names and comments, grepping code, and even 
asking a teammate what they were thinking when the wrote it. Humans fundamentally have a limited amount of memory and tend to 
fatigue. Thus, they often take mental short-cuts and make assumptions based on limited syntactic signal and prior experience. These tricks 
are usually safe and eliminate substantial amounts of work. However, when there are subtle, unusual, or special context specific behaviors, these 
reasoning shortcuts will miss them and result in erroneous behaviors. These failures range from corner case bugs with indeterminate behaviors like 
unstable sorts~\cite{jssort}, to large scale outages cased by oversights in recovery logic~\cite{failurefailure}, or industry-wide security 
issues from assuming that a \cf{log} statement would never access a remote server and execute un-trusted code~\cite{log4j}. In hindsight 
all of these issues are ``obvious'' but as humans we are all subject to fatigue, oversights, and heuristic biases that let assumptions derived 
from surface syntax and implicit expectations lead to missing critical behaviors. 

\noindent
\paragraph{\bf Symbolic Analysis Tooling:} Symbolic analyses in many ways are on the opposite end of the spectrum from a human developer 
in terms of their reasoning strengths and weaknesses. They do not understand anything apart from the code itself as written and do not bring 
any preconceived notions about what it might do. Thus, these agents will never overlook a subtle corner case or possibility. However, this 
attention to specifics and lack of intuitive understanding often leads to them getting ``lost in the details'' and being unable to scale to realistic 
codebases where they must deal with non-determinism, loop/inductive invariant generation, aliasing, and more. In practice, to be feasible, they 
are often simplified substantially, at a cost of imprecision, which generally leads to issues with false positives/negatives and reduces the value
they can provide.

\noindent
\paragraph{\bf AI Agents:} Large-Language-Model (LLM) agents~\cite{gpt4} are an interesting addition to this spectrum. They can perform heuristic resesoning like humans but they 
do not get tired and can hold huge amounts of context while performing a task currently up to 32k tokens (``words''). However, these models are strongly 
syntax based and, as a result, if there are important \emph{semantic} properties of the code that are not \emph{syntactically} visible then the agent 
must fall back to likelihoods learned in the training data. These features make AI agents remarkably capable when performing certain coding 
tasks. However, these properties also lead them to making mistakes that can be quite perplexing. Current LLM's (like GPT-4) 
are prone to making-up and using API's that sound plausible but that don't actually exist or using API's in ways that defy common sense -- for 
example picking a hard-coded time to schedule a meeting instead of \emph{correctly} querying a schedule availability API to find a mutually open 
slot. The propensity to make these types of mistakes is compounded by the inability of these AI agents to leverage tools to validate assumptions 
in code they generate. Thus, just like human developers, improvements in symbolic reasoning based tools can lead to immediate improvements 
in the performance of statistically based AI agents as well.

\subsection{Sources of Complexity}
\label{sec:complexitysources}
Given the capabilities of the agents we want to support, humans, symbolic tools, and AI systems, we next look at seven commonly appearing 
features in modern languages that impede their reasoning.

\noindent
\paragraph{\bf Mutable State and Frames:}
Mutable state is a deeply complicated concept to model and reason about. The introduction of mutability into 
a programming language destroys the ability to reason about the application in a \emph{monotone}~\cite{ppa} manner which forces the programmer 
(and any analysis tools) to identify which facts remain true after an operation and which are invalidated. The ability for mutable code to affect the state of the 
application via both return values and side affects on arguments (or other global state) also introduces the need to reason about the 
\emph{logical frame}~\cite{frameproblem,seplogic} of every operation. 

\noindent
\paragraph{\bf Implicit Behaviors:} 
Behaviors that are not explicit in the syntax of the code create opportunities for misunderstanding, oversights, and increase the cognitive load when 
reasoning. Examples of this include implicit nullability, the ``billion dollar mistake'', unchecked exceptions, and APIs with undeclared implicit error behaviors. A 
common example is the \texttt{min} operation in C\# \vs Python. In C\# the \texttt{min} operator return is declared as \texttt{Option<T>} so that 
the possibility of a invalid result is syntactically explicit in the declaration. In Python one must look into the API docs to find the comment that the 
\texttt{min} function throws an error on the empty list.

\noindent
\paragraph{\bf Hidden Semantics:} 
Similar to implicit behaviors, many languages have implicit rules that affect program behaviors. These include implicit type coercion, numeric truncation, 
and flow-sensitive or other advanced of type inference algorithms. These increase the cognitive load to understand a block of code as they require additional work to simulate 
a non-trivial hidden algorithm to extract implicit program semantics and create opportunities for simple oversights to create unexpected outcomes.

\noindent
\paragraph{\bf Loops, Recursion, and Invariants:} Loops and recursion represent a fundamental challenge to reasoning as the code describes the effects of a 
single step but understanding the full construct requires \emph{generalization} to a quantified property over a set of values. Invariants~\cite{finvariants,hinvariants} 
provide the needed connection but a generalized technique for their computation is, of course, impossible in general and has proved elusive even for restricted 
applications. 

\noindent
\paragraph{\bf Indeterminate Behaviors:} Indeterminate behaviors, including undefined, under specified, or non-deterministic or environmental behavior, require a programmer 
or analysis tool to reason about and account for all possible outcomes. While truly undefined behavior, \eg uninitialized variables, has disappeared from most 
languages there is a large class of under-specified behavior, \eg sort stability, map/dictionary enumeration order, etc., that remains. These increase the complexity 
of the development process and, as time goes on, are slowly being seen as liabilities that should be removed~\cite{jssort}. Less obviously the inclusion of non-deterministic 
and/or environmental interaction results in code that cannot be reliably tested (flakey tests), behaves differently for non-obvious reasons, and frequently mixes failure logic 
widely through a codebase. 

\noindent
\paragraph{\bf Data Invariant Violations:}
Programming languages generally provide operators for accessing, and in imperative languages updating, individual elements in arrays/tuples or fields in objects/records. 
The fact that these accessors/updaters operate on an individual elementwise basis results in programmers updating the state of an object over multiple 
steps, or manually exploding an object before creating an updated copy, during this span invariants which normally hold are temporarily invalidated before 
being restored. During these intervals the number of details that must be tracked and restored can increase drastically increasing opportunities for mistakes 
and oversights to occur. 

\noindent
\paragraph{\bf Equality and Aliasing:} Programming languages live at the boundary of mathematics and engineering. Although language semantics are 
formulated as a mathematical concept there are common cases, \eg reference equality, pass by-value \vs by-reference, or evaluation orders, that expose and 
favor one particular hardware substrate, generally a \emph{Von Neumann} architecture, either intentionally for performance or accidentally by 
habit or history. While seemingly minor these choices have a major impact on comprehensibility -- merely exposing reference equality pulls in the complexity of reasoning about aliasing relations and greatly complicates compilation on other architectures.

\section{\bosque Design Philosophy}
\label{sec:philosophy}
The unique value proposition of \bosque is not based on a single big new feature, or even a number of smaller novel features, instead the novelty comes from a holistic 
process of simplification and feature selection with a single focus toward what will simplify reasoning about code. The first part of this process is eliminating, or minimizing, 
the sources of complexity identified in \autoref{sec:complexity}. This process primarily involves the design of the \bosqueir in \autoref{sec:bsqir}. The second part of the 
process centers around how to connect this core IR to a user friendly surface syntax, and, how to make this syntax amenable to human and LLM agent understanding. 
This involves the development of the language structure and the introduction of a variety of intentional but subtle features that fade into the background and lead 
developers onto a path of producing clean code without effort or conscious thought.

The idea of a \emph{solver integrated} or aware language has been explored in several contexts -- examples include Rosette~\cite{rossette}, why3~\cite{why3}, Dafny~\cite{dafny}, and Ivy~\cite{ivy}. 
The design of \bosqueir is unique in that it does not attempt to expose the solver, including limitations it has, to the developer. Instead it is designed to naturally encourage 
and simplify the mapping of code written by a non-expert developer (one who is not an expert in logics or provers) into one of many possible reasoning frameworks such as SMT solvers, 
deductive logical engines, abstract interpretation, and simple dataflow solvers. 

At the core of \bosqueir is a let-based functional language. Although there are many functional languages 
that have key properties we want in the IR, in their attempts to also be general purpose languages or due to historical accident, they also end up with features like mutable cells, 
visible aliasing, lazy evaluation, non-local effects and flow, or indeterminate behaviors in their specifications. Thus, the \bosqueir can be seen as a highly \emph{regularized} functional 
language that is fully aligned with the goal of eliminating the complexities identified in \autoref{sec:complexity}.

In \autoref{sec:bsqsrc} this paper shows how a selection of language features and syntactic structures can connect a regularized IR like the \bosqueir to a block-structured (and imperative looking) 
source language that developers are comfortable with. As described in \autoref{sec:bsqsrc} we can use syntactic sugar to support commonly used and easily understood programming constructs 
like block-structured code, reassignment of variables, updates through (shallow) references, early returns, and object-oriented data types. These constructs are heavily used and well understood 
features in modern software engineering so supporting them allows developers to easily express their intents in a natural way. However, by our careful construction, these features can in fact be 
compiled directly down into the core \bosqueir representation. Thus giving us the best of both worlds, where developers can express their domain and business logic in an intuitive syntactic form 
and, unlike in a standard imperative or object-oriented language, the underlying semantics are the well-behaved \bosqueir structures.

\section{\bosqueir Intermediate Representation}
\label{sec:bsqir}

Given the sources of complexity identified in \autoref{sec:complexity}, the first task is to construct an 
intermediate language (IR) that eliminates or reduces the impact of each of them. As many of the 
issues are tied to mutability and loops, the design process starts with a strict functional language core. 
This core uses classic \cf{let} bindings, \cf{if-then-else} conditionals, a special assert expression, and 
a restricted definition of equality.

%

\subsection{Primitive Types and Values}
The \bosqueir language provides a standard assortment of primitive types and values including a special \cf{none} 
value (\cf{None} type), a \cf{Bool}, \cf{Nat} and \cf{Int} numbers, safe \cf{BigNat} and \cf{BigInt} numbers, along 
with \cf{Float}, \cf{Decimal}, and \cf{Rational} numbers. The \bosqueir \cf{String} type represents immutable 
unicode string values (a special \cf{ASCIIString} is also provided). The language includes support for commonly 
used types/values in modern cloud applications like ISO/Plain times, SHA hashcodes, UUIDs, timestamps, and 
other miscellany. 

\subsection{Self Describing Types and Values}
Structural \emph{Tuple} and \emph{Record} types provide standard forms of self describing types. Structural types 
do not allow subtyping (\ie not covariant). This allows for an analysis to know, and precisely encode, the types of 
values in a structural type. This representation results in substantial simplifications when reasoning about 
operations on these values as lookups are always statically resolved. Examples of these types include:

\begin{lstlisting}[language=Bosque]
[Int, String] //a tuple type
[5i, "ok"] //a tuple of type [Int, String]

{p1: Int, p2: String} //a record type
{p1=5i, p2="ok"} //a record of type {p1: Int, p2: String}
\end{lstlisting}

The \bosqueir representation also supports self describing ad-hoc \emph{union} types. The sub-typing 
relation on union types is defined as being a subtype of any enumerated type in the union.

\begin{lstlisting}[language=Bosque]
Int | None //a union type
if true then 5i else none //expression of type Int | None
\end{lstlisting} 

\subsection{Nominal Types}
\bosqueir supports a nominal type system that allows the construction of object-oriented style type inheritance 
structures. Abstract \cf{concepts} provide a means for inheritance and multi-inheritance via \emph{conjunction}. 

In the following example there are two concepts declared, \cf{WithName} and \cf{Greeting}, that define a field 
and abstract method respectively. The concrete entity \cf{GenericGreeting} is declared to provide the \cf{Greeting} 
concept and, implements the required \cf{sayHello} method. The \cf{NamedGreeting} provides both the \cf{WithName} 
concept, so it inherits the field \cf{name} and the \cf{Greeting} concept. Using the \cf{name} field it implements the 
\cf{sayHello} method to return a customized string with the name information.

\begin{lstlisting}[language=Bosque]
concept WithName {
  field name: String;
}

concept Greeting {
  abstract method sayHello(): String;
}

entity GenericGreeting provides Greeting {
    override method sayHello(): String {
        return "hello world";
    }
}

entity NamedGreeting provides WithName, Greeting {
    override method sayHello(): String {
        return String::concat("hello ", this.name);
    }
}
\end{lstlisting}

In \bosqueir, nominal subtyping is based on the transitive closure of the provides relation. In this example the entity 
\cf{GenericGreeting} is a subtype of just \cf{Greeting} while the entity \cf{NamedGreeting} is a subtype of both 
\cf{WithName} and \cf{Greeting}.

The nominal type system differs from many object-oriented type systems in that the \emph{concepts} are always 
abstract and can never be concretely instantiated while \emph{entity} (class) types are always concrete and can 
never be subclassed. This prevents confusion in type checks where an \cf{instance of} test may be intended to 
check for an exact type but actually includes all subtypes as well. More generally this enables us to easily determine 
if a given operation has a single concrete behavior or may be dynamic in nature.

{\small
\begin{figure*}
\centering
\begin{eqnarray*}
\mbox{ListStructure} & \bnfas & \cf{List<T>}\{e_1, \ldots, e_j\} \bnfalt \cf{slice}(l, i, j) \bnfalt \cf{concat}(l_1, l_2) \ldots \\
\mbox{ListAccess} & \bnfas & \cf{size}(l) \bnfalt \cf{get}(l, n) \bnfalt \ldots \\
\mbox{ListCompute} & \bnfas & \cf{map<fn>}(l) \bnfalt \cf{filter<p>}(l) \bnfalt \bnfalt \cf{join<p>}(l_1, l_2) \bnfalt \ldots \\
\mbox{ListPred} & \bnfas & \cf{has<p>}(l) \bnfalt \cf{find<p>}(l) \bnfalt \cf{count<p>}(l) \bnfalt \ldots \\
\mbox{ListIterate} & \bnfas & \cf{sum}(l) \bnfalt \cf{reduce<fn>}(l) \bnfalt \ldots 
\end{eqnarray*}
\caption{\bosqueir \cf{List<T>} Operations}
\label{fig:list}
\end{figure*}
}

\subsection{Typedecls}
\label{sec:typedecls}
Primitive datatypes are a frequent source of \emph{implicit} non-syntactic information in code. Strings and 
integers are often used to represent semantically distinct concepts, like a zipcode or a temperature, just via 
convention and use. These uses make certain classes of bugs possible, like adding a temperature in Fahrenheit 
to one in Celsius, and hide important information in implicit channels. Sometimes this design is motivated by 
performance concerns, \eg space/boxing overheads, but is often just an issue of simply wanting to avoid the 
effort to declare a new datatype and implement all the associated methods (particularly for numeric types). 

The \bosqueir representation provides explicit support for these cases via a \cf{StringOf} type and \cf{typedecl}s 
that manage the new-type creation, ensuring that the type representation is isomorphic to the underlying type, 
generation of the standard operators, and allow customization of the underlying value with user-defined invariants. 

The example below shows the declaration of a typed string \cf{Zipcode} using a regex and the parametric 
\cf{StringOf} type. These structured strings are frequently used in modern web applications -- OpenAPI~\cite{openapi} and 
Microsoft TypeSpec~\cite{typespec} both natively support similar constructs. As shown in the example this allows us 
to declare typed string literals, check they satisfy the constraints on construction, and are used in type-safe 
ways. 

\begin{lstlisting}[language=Bosque]
typedecl ZipcodeValidator = /[0-9]{5}(-[0-9]{4})?/;
typedecl Zipcode = StringOf<ZipcodeValidator>;

typedecl Celsius = Float;

//create a literal zipcode string
"40502"Zipcode //Ok
"ABC"Zipcode   //Type Error

function isNYZipcode(zc: Zipcode): Bool {...}

isNYZipcode("40502"Zipcode) //Ok returns false
isNYZipcode("40502")        //Error arg is a String

let temp = 10_Celsius + 1_Celsius  //Type safe operation
\end{lstlisting}

The \cf{typedecl} type constructor also supports creating new types for any primitive type such as Ints, 
Floats, UUIDs, etc. In the example a new type for the temperature \cf{Celsius} is created and, all well-typed, 
operations are permitted on the new numeric types.

\subsection{Key Types and Equality}
\label{sec:keytypes}
Equality is a multifaceted concept in programming~\cite{lefthandequals} and ensuring consistent behavior across 
the many areas it surfaces in a modern programming language such as \lstinline{==}, \lstinline{.equals}, 
\lstinline{Set.has}, and \lstinline{List.sort}, is source of subtle bugs~\cite{findbugs}. This complexity further 
manifests when equality can involve referential identity which introduces issue of needing to model aliasing 
relations on values, in addition to their structural data, in order to understand the equality relation.

To avoid these behavioral complications, and the need to model aliasing, the \bosqueir language is 
\emph{referentially transparent}. The only values which may be compared for equality are specific 
primitive values including \emph{none}, booleans, primitive integral numbers, strings, and typedecls of 
these primitive types. In conjunction with the immutability of the values (\autoref{sec:operations}) this 
ensures that \bosqueir code is referentially transparent and functions do not need to use \emph{frame rules}~\cite{frameproblem}. 

\subsection{Operations}
\label{sec:operations}
The expression language for \bosqueir is designed explicitly 
with the needs of symbolic reasoning systems in mind. The language allows recursion but most code is expected to 
use a rich language of higher-order functor libraries (see \autoref{fig:list}). Otherwise, the \bosqueir expression 
semantics are constructed to ensure referential transparency, that all expressions are side-effect free, that 
any expression deterministically evaluates to unique result value and, with the inclusion of an error-result, that 
they are total as well. 

Primitive expressions include special constants like \cf{true}, \cf{false}, \cf{none}, literal integral \cf{i} values or literal 
float \cf{f} values, literal strings \cf{s}, and variables (either local, global, or argument).
\bosqueir has the standard assortment of numeric and logical \emph{Operators}.
The \cf{constructor} operations are all simple and explicit operations for tuples and records. The constructors 
for nominal entity types takes the type name + the full list of values to initialize the fields with. 
Deconstructing compound values is done with a standard indexing for tuples ($e.i$ where $i$ is a constant), 
or named property/field accessor ($e.p$ and $e.f$ respectively). 

The type manipulation operators include standard \cf{is} and \cf{as} operations to test value types (or subtype) and 
to perform (checked) casts on them. The \cf{inject} operation is used to convert a primitive value into a matching \cf{typdecl} 
type and the and \cf{extract} operation extracts primitive values. The assert operation provides a way to explicitly generate 
an error value (for user defined errors) in addition to implicitly defined error sources, like cast failures, out-of-bounds indexing, \etc.

\subsection{Collection Functors}
\label{sec:functors}
Containers and operations on them play a major role in most programs and these semantics inherently involve reasoning 
over the container contents. Instead of introducing a primitive looping construct, and attempting to tackle the loop invariant 
generation problem, the \bosqueir language includes a rich set of container datatypes and functor based operations 
on them. In practice these operations, and parameterizeable functors, are sufficient to cover the majority of 
iterative operations~\cite{loopmining}. 

\autoref{fig:list} shows a sample of the operations that are provided for processing lists (maps, sets, stacks, and queues are also 
supported). These operations provide a structured way to process the values in collections. Most of these operations are 
familiar from libraries such as LINQ~\cite{linq}, Java Streams~\cite{javastreams}, or lodash~\cite{lodash}. One unique restriction 
on them is that the lambda parameters are passed as parametric arguments which specialize each functor. This preserves the 
first-order nature of the language, since we do not need to pass function types, and as we see later, can be ensured with a 
simple syntactic restriction on how lambdas can be used in the source language.

\section{\bosque Source Representation}
\label{sec:bsqsrc}

The construction of the \bosqueir language, as described in \autoref{sec:bsqir}, is free of many features that 
complicate reasoning \wrt application behavior. However, it is not particularly human (or LLM AI) friendly and  
does not address many problems identified in the context of implicit behaviors or hidden semantics. At a high-level the \bosque language 
is designed to be easily accessible to a developer coming from TypeScript, C\#, or Java while mapping well to the restricted \bosqueir representation. 
\bosque uses a block-based structure, allows reassignment of variables, has an extensive set of control-flow operators, object-oriented language 
features, and other language features designed to simplify common programming idioms. 

This section describes the specific features of the \bosque source language exposed to developers that are related to our key objective, supporting reasoning. 
\autoref{fig:itree} is a partial implementation of a binary-tree in \bosque that contains many of these features. It uses the 
extended Algebraic Data Type (ADT) declaration form (\autoref{sec:enahncedtypes}) to declare the various types associated with a tree, early returns and other block structured control flow 
including variable re-assignment (\autoref{sec:blockrebind}), explicit flow-typing and binding (\$) as described in \autoref{sec:flowbind}, explicit recursion tags to denote 
recursive call paths (\autoref{sec:explicitrec}), and the use of data invariants (\autoref{sec:datavalidation}). Later we will describe how \texttt{ref} methods work and show
how they allow simulated in place (\autoref{sec:refmethods}) updates.

\begin{figure}
\begin{lstlisting}[language=Bosque]
datatype ITree using {
    size: Nat
} of
  Nil {}
  | Leaf { v: Int }
  | Node {
      invariant test size == l.size + r.size + 1n;
      
      field v: Int; 
      field l: ITree; 
      field r: ITree;
  }
& {
  const empty = Nil{0n};
    
  method isEmpty() {
    return this?<Nill>;
  }
    
  recursive method has(x: Int): Bool {
    match(this) {
      Nill => return false;
      | Leaf => return $.v == x;
      | Node => {
        if($.v == x) {
          return true;
        }
                
        var tchild: ITree;
        if(x < $.v) {
          tchild = $.l;
        } 
        else {
          tchild = $.r;
        }
        
        return tchild.has[recursive](x);
      }
    }
  }
} 
\end{lstlisting}
\vspace{-4mm}
\caption{Bosque Binary-Tree Example}
\label{fig:itree}
\end{figure}

\subsection{Enhanced Algebraic Data Types}
\label{sec:enahncedtypes}
The type system that is presented to the surface user of the \bosque language closely mirrors the type system of the core \bosqueir language. The 
major ergonomic addition in the \bosque source language is the provision of syntax for creating algebraic-data-types.
This syntax allows for the compact subtyping, as provided by most ADT implementations, along with inheritance of fields and definition/override of methods, invariants, \etc 
These features make the ADT syntax more useful and practical. 

In the binary-tree example these features allow us to declare a field \cf{size} which is inherited by each of the cases, \cf{Nil}, \cf{Leaf}, and \cf{Node}. The simple cases of 
\cf{Nil} and \cf{Leaf} are similar to the traditional ADT style where only the public fields are specified. When, inevitably the data types get more complex as in the case of the
\cf{Node} type, \bosque supports expanding the declaration with \cf{invariants} and other type members. To simulate the desirable encapsulation properties of object-oriented 
languages this declaration allows a tailing block that can include any member type declarations. These are all defined in the scope of the enclosing ADT type, in our example 
the \cf{ITree} type. The example includes declarations of a member constant, \cf{empty}, and three member methods. This shorthand provides a compact way to group 
logically related subtypes and operations in addition to avoiding the need for mostly trivial class declarations (in our examples it would be $1$ abstract class plus $3$ concrete 
classes in a language like Java or TypeScript).

\subsection{Explicit Flow Typing and Binding}
\label{sec:flowbind}
A key principle in the design of the \bosque surface language is to ensure that, as much as possible, the behavior of an expression is explicit in the syntax. One area 
where behavior is often implicitly encoded is type flow and automatic inference/coercion. Flow based typing is a very convenient feature but requires a user of the 
language to know the flow-rules~\cite{typescript,rust} of the checker, and to run the algorithm, to understand type properties~\cite{rust,rustflow}. As a balance 
between the convenience of implicit flow-typing and the simplicity of explicit type coercion, \bosque provides specialized explicit flow typing expressions and binding. 

The explicit flow type operations shown in \autoref{fig:fexpressions} are designed to cover common idiomatic type tests (FlowSpecial), support standard is/subtype 
checks (FlowType), and enable flow type inference when testing against constant values (FlowEQ). The ability to negate these tests allows them to be used easily in 
control-flow contexts as well. These operations can be used for \emph{is-testing} using the \cf{?} operator or \emph{as-casting} using the \cf{@} operator. This 
explicit operations have the benefit of clearly indicating in the source which variables and type constraints are of relevance, it is applicable to arbitrary expressions as 
opposed to only variables or some limited forms, and it ensures that they compiler type-checker does not experience compounding growth in managing flow-type 
information (\eg exponential path-sensitivity) or arbitrary limits on what inferences\footnote{From a engineering perspective it is also removes the very painful problem 
of later changes to these type-inference heuristics introducing (subtle) bugs into an application!} are made (\eg discard on path joins, only top-level variables, ...).

{\small
\begin{figure}
\begin{eqnarray*}
\mbox{FlowSpecial} & \bnfas & \cf{none} \bnfalt \cf{some} \bnfalt \cf{ok} \bnfalt \cf{err} \bnfalt \cf{result} \\
\mbox{FlowType} & \bnfas & \cf{<} \emph{Type} \cf{>} \\
\mbox{FlowEQ} & \bnfas & \cf{[}\emph{literal}\cf{]} \\ 
\mbox{FlowAlt} & \bnfas & \emph{FlowSpecial} \bnfalt \emph{FlowType} \bnfalt \emph{FlowEQ}\\
\mbox{FlowOp} & \bnfas & \emph{FlowAlt} \bnfalt \cf{!}\emph{FlowAlt}  
\end{eqnarray*}
\caption{\bosque Flow Typing and Binding}
\label{fig:fexpressions}
\end{figure}
}

Examples of these operations for type testing and coercion expressions include:
\begin{lstlisting}[language=Bosque]
x?some    //test if x is a subtype of Some
x@some    //coerce of x to a Some type or fail
x?!<Nat>  //test if x is not a Nat
x@[5i]    //test if x is 5 and coerce to Int type
foo(x).f?<Bar> //test if arbitrary expression is type Bar
\end{lstlisting}

The typing tests also enable special narrowing statements in blocks -- rebinding the type of the variable. In this case narrowing the type of \cf{x}, 
which is initially declared as a \cf{Nat | none}, to be a \cf{Nat} (and raising a runtime error if it is not). This explicit narrowing, as opposed to an 
implicit flow-sensitive analysis around say, \cf{assert(x != none)}, enables us to explicitly encode the intent in the source so that it is explicit and 
obvious to humans or AI agents.
\begin{lstlisting}[language=Bosque]
let x: Nat? = ...;
let e = x + 2n; //type error
...
//check that x is a Nat (error otherwise) and narrow
x@<Nat>;
...
let y = x + 2n; //ok
\end{lstlisting}

The typing tests are also used in control flow expressions and statements to provide bindings in branches of control flow structures. \bosque provides the special 
\cf{\$} variable that is automatically bound to the narrowed results of the conditional guard. Consider the following example where a variable may be \cf{Nat | none}. 
With the flow-type check \cf{if none (x)} we know that on the false branch the type of \cf{x} must be a \cf{Nat}. Instead of leaving this information implicit 
and simply reusing the variable \cf{x} \bosque code leaves the type of \cf{x} unchanged and introduces the new binder \cf{\$} with the same value but the 
type restricted based on the test.

\begin{lstlisting}[language=Bosque]
let x: Nat? = ...;
if none (x) {
    return 0n;
}
else {
    //Binder for $ to value of x (with type Nat)
    return $ + 10n;
}
\end{lstlisting}

Finally, these \emph{FlowOp} typing tests are also used to handle early return statements in blocks. In the following example the \cf{??} operator syntactically 
reduces to \cf{if none (foo(...)) \{return none;\} let x = foo(...)@!none; } while the \cf{@@} operator reduces to the code \cf{if !<Nat>(x) \{return x@Int;\} x@<Nat>; }. 
Thus, this syntax allows a developer to succinctly express these common check-and-early-return idioms in an explicit way and without filling the code with conditional 
flows that obscure the underlying algorithm.
\begin{lstlisting}[language=Bosque]
function bar(k: Int): Nat | Int | None {
    //test/narrow call result -- return none if fails 
    let x: Nat | Int = foo(...) ?? !none;
    let e = x + 2n; //type error
    ...
    //check/narrow type of x -- return Int otherwise
    x @@ <Nat>;
    return x + 2n; //ok
}
\end{lstlisting}

These operators provide shorthand notation for common idioms~\cite{idioms} that developers use \emph{and} ensure that they are 
syntactically surfaced in the code. This both simplifies writing the code, reduces the opportunities for developers to miss important 
behaviors, and provides AI agents with explicit representations.

\subsection{Lambdas}
Higher order functions are a powerful programming mechanism that can be used to great effect to simplify code. However, their 
use can also lead to difficult to reason about flows and inscrutable behaviors. Empirical work on the use of lambdas in large codebases~\cite{jlambda} 
provides a compelling quantitative result that matches the anecdotal intuition. In particular the vast majority of use of lambdas 
are in direct positions ($87.5\%$ are passed directly as arguments in calls) and very few examples of classic functional programming 
techniques, like currying,  were seen.

With this in mind the \bosque language provides a syntactically restricted version of lambdas where they:
\begin{enumerate}
\item Cannot be stored in local variables or returned as function/method results
\item Can capture arguments but \emph{cannot} modify their values
\end{enumerate}

\bosque lambdas are also split into two categories. Predicates are denoted with the \cf{pred} keyword, which must return a \cf{Bool} result, and other functions 
that may return any type with the \cf{fn} keyword.\\

\begin{lstlisting}[language=Bosque]
//takes a predicate
l.allOf(pred(x) => x >= 0i)

//takes a function
l.map<Int>(fn(x) => x + 1i)
\end{lstlisting}

In combination these restrictions and conventions provide the vast majority of benefit that most developers get from lambda style expressions while prohibiting 
their more problematic uses. These restrictions also ensure that we can syntactically convert code using lambdas to a de-functionalized form where the functions 
are reduced to direct parametric specializations as needed by the \bosqueir language (\autoref{sec:functors}).

\subsection{Block Structure with Re-bindable Variables}
\label{sec:blockrebind}
Interviews and conversations with developers, ranging from fortune 100 companies to 3-person startups, suggested a strong preference for block-scoped 
flows and mutable variables over a classic-functional let-bound expression tree (which \bosqueir is). In order to unify these two representations we leveraged 
the loop-freedom of the source language to convert multiple-assignments and convergent-dataflow into \emph{dynamic single assignment} form~\cite{dsaform}. 
Consider the code:
\begin{lstlisting}[language=Bosque]
function abs(x: Int): Int {
    var y = x;
    
    if (y < 0) {
        y = -y;
    }

    return y;
}
\end{lstlisting}

This function shows the use of multiple updates to the same variable and block structured conditional flows. We distinguish between variables, \cf{let}, 
that are fixed and those, \cf{var}, that can be updated. This ability to set/update a variable as a body executes simplifies a variety of common coding 
patterns.

Since the language is loop free, it can be easily converted to a SSA form~\cite{ssa}, and the loop freedom also ensures that any assignment is also 
a single assignment dynamically! A simple treeification pass, with the introduction of continuation functions if needed, also eliminates all DAG control 
flow. Thus, our block structured flows and updated variables becomes the following let-bound expression:
\begin{lstlisting}[language=Bosque]
function abs(x: Int): Int {
    let (y = x) in
    if y < 0 then -y else y
}
\end{lstlisting}

Similar transformations allow \bosque to support many other familiar block structured flow features. The running binary-tree example has an implementation 
utilizes these heavily, including early returns, merging control flows, creating and later assigning variables (the \cf{tchild} variable) and checking for any possible 
uninitialized uses. These constructs are heavily used and well understood features in modern software engineering so supporting them allows developers to 
easily express their intents in a natural way.

\subsection{Constructrors and Bulk Operators}
\label{sec:consbulk}
Even with immutable objects there can be subtle challenges with constructor semantics and implementing 
operations which create a copy of a value with updates to some subset of the contained values. Constructor bodies where fields are initialized sequentially with, 
potentially other computation mixed in, can lead to issues where methods are invoked on partially initialized values. Updates to objects are often implemented 
by copying fields/properties individually while replacing the some subset with new values. These issues can lead to both subtle bugs during initial 
coding \emph{and} also make it difficult to update data representations when refactoring code or adding a new feature at some later date.

Consider the code shown below. The use of atomic constructors prevents the partially initialized object problem~\cite{oinv1,oinv2,joed} 
when constructing the \cf{Baz} object. This example also shows how the \emph{bulk algebraic} operations simplify the update/copy of the \cf{Baz} object 
and eliminate the problem of temporarily violated invariants during a series of single field updates\footnote{Fields can be declared as private, in which case 
these \emph{raw} operations will not be visible outside the type declaration.}. \bosque provides explicit support for 
data invariants, \eg in the \cf{Node} entity in our example) and described more below, these are automatically checked on both initial construction and any 
updates.

\begin{lstlisting}[language=Bosque]
concept Bar { 
	field f: Int; 
}

entity Baz provides Bar {
	field g: Int;
	field h: Bool;
}

//create a Baz value with all fields initialized
var x = Baz{1, 2, true};

//copy of x with f updated to 3 and h to false
var y = x.{f=3, h=false};
\end{lstlisting}

To make these operations more useful they also handle the common case of wanting to update a field with a function of that field value. This is done by 
extending the range of where \emph{binders} can be used and, for each field updated, creating a variable that is scoped to the update expression and 
that is bound to the original value of the field. In the code below this allows us to increment the value of the field \cf{f} by $3$ and then assign the new 
value in the update.
\begin{lstlisting}[language=Bosque]
//copy of x with f incremented by 3
var y = x.{f=$f + 3};
\end{lstlisting}

\subsection{Data Validation}
\label{sec:datavalidation}
Program logic checks are fundamental to quickly catching bugs and make implicit assumptions explicit in program. Thus, the \bosque surface language provides 
a rich set of validation features to support a broad range of uses including dynamic checking, static validation, optimization, and documentation for human (and AI) developers. 
Including them as first class components of the language provides several advantages over libraries or 3rd party implementations. 

From a reasoning perspective having builtin validation operators ensures that every application will have the same look and semantics around assertions. It also 
gives the compiler and runtime direct knowledge of these special operations for example an optimizer can move them off the hot path aggressively, do short circuit 
evaluation of any message or line number computations, and, if they are disabled, can easily remove all of the dead code. In \bosque we use this awareness
along with a new level, \cf{safety}, that we ensure is always checked -- \eg the compiler is never allowed~\cite{optbug} to optimize it out!

\noindent
\paragraph{Assert/Pre/Post:} The first form of validation is a classic conditional assert statement that can be used to place ad-hoc checks in a block of code. 
\bosque also supports pre/post conditions on functions and methods. These features allow developers to insert, explicit, information on expectations/assumptions 
for any bit of code or invocation.

\noindent
\paragraph{Invariants and Validates:} The ability to explicitly state data invariants is one of the most powerful validation features in the \bosque language. These invariants 
allow a developer to state a property in a single location -- this property ensures that at every creation site it must be preserved and provides a guarantee for 
every use of the type in a program. An example invariant is seen in the binary-tree example.

Ingesting data from external sources, such as command line args, network data, file reads, \etc, is a critical task. Writing code to validate data, 
even structured data in JSON or XML form, is a tedious and error prone task~\cite{restler}. Errors in this code are amplified as they open opportunities for external, 
and potentially malicious, sources to directly interact with the application. 

By default \bosque checks all active invariants whenever an value is constructed. These are generally not as extensive as would be needed to fully validate untrusted inputs. 
Thus, \bosque provides a \cf{validate} keyword that allows the specification of checks that must be run on eternal inputs. When compiling \bosque to executable code 
these \cf{validate} checks are combined with the \cf{invariants} in a special function that the host can use to create values from untrusted external data sources.

\autoref{fig:saleinfo} shows code from a sample trading application provided by Morgan Stanley that was ported to \bosque (see \autoref{sec:casestudies}). 
In this code there are several invariants and external validations on the \cf{SaleInfo} type. The check \cf{available >= 0I} is performed everytime an 
\cf{SaleInfo} value is created. The invariant \cf{startAvailable >= 0I} is marked as \cf{test} so it is only enabled when running the code in a test build. 

The two \cf{validate} checks are too expensive to run on every internal operation \emph{but} if we received a JSON value encoding this info from, say, a HTTP 
request from a 3rd party we definitely want to check that the data is well formed and consistent with our requirements (we can also use these for static verification).  

\noindent
\paragraph{Levels:} The \cf{validate} feature is a special, and very important, case of the general problem of balancing checking useful properties 
against the cost of running these checks. To support the ability to utilize these specification features without concern about how they will impact 
the performance end-users experience \bosque allows any use of a validation annotation to be pre-fixed with a level, \cf{spec}, \cf{debug}, \cf{test}, 
\cf{release}, or \cf{safety}. 

The \cf{debug}, \cf{test}, \cf{release} levels are useful for controlling which checks are run dynamically under which 
conditions. The \cf{spec} level is useful for checks which would always be infeasible to check at runtime but which are useful for documentation, 
static analysis tools, and sampling based checking if a program is run in debug build mode. The \cf{safety} level is for checks that a developer 
wants to run, even if they can be proven to always hold! This counter-intuitive feature is to ensure that a compiler will never eliminate 
tests that are critical to data integrity and may still be possible due to hardware or other failure modes.

\begin{figure}
\begin{lstlisting}[language=Bosque]
entity SaleOrder {
    field id: StringOf<ValidID>;
    field quantity: BigInt;
}

entity SaleInfo {
    field available: BigInt;
    field startAvailable: BigInt;
    field orders: List<SaleOrder>;
    
    //check sanity on every operation
    invariant available >= 0I;
    invariant test startAvailable >= 0I;
    
    //too expensive on every change 
    //but *must* check on untrusted inputs
    validate orders.unique(pred(a, b) => a.id !== b.id);
    validate startAvailable - orders.sumOf<USD>(fn(a) => a.quantity) == available;
}
\end{lstlisting}
\vspace{-4mm}
\caption{Declarations from Sample Trading App}
\label{fig:saleinfo}
\end{figure}

\subsection{By-Ref Methods}
\label{sec:refmethods}
As very common task is sequentially processing data with an environment of some sort. This can be clumsy 
to do manually, requiring manual packing and unpacking of env/value results, and the common functional solution of introducing monadic features clashes 
with our desire to keep behavior syntactically explicit. So, \bosque introduces the concept of \cf{ref} methods. These are explicitly tagged at both def 
and call sites and manage the update of the receiver variable with the new state automatically.

In the following example the \cf{Counter} is initialized to $0n$ and at each \cf{ref} method invoke the receiver variable is updated with the result of the \cf{this} value in
the called method as well as assigning the result value. The statement \cf{this.\{ctr = \$ctr + 1n\};} updates the value of \cf{this} with the new \cf{ctr} value. Any calls to the 
\cf{generateNextID} are required to be top-level (not nested in other expressions) and annotated with a \cf{ref} attribute. If either of these conditions are not satisfied 
the type-checker will reject the code.

\begin{lstlisting}[language=Bosque]
entity Counter {
    field ctr: Nat;

    function create(): Counter {
        return Counter{0n};
    }

    method ref generateNextID(): Nat {
        let id = this.ctr;
        this.{ctr = $ctr + 1n};

        return id;
    }
} 

var ctr = Counter::create();  //create a Counter 

//id1 is 0 -- ctr is updated
let id1 = ref ctr.generateNextID();

//id2 is 1 -- ctr is updated again
let id2 = ref ctr.generateNextID();
\end{lstlisting}

\subsection{Recursion}
\label{sec:explicitrec}
Complex recursive control flows obfuscate the intent and hinder automated analysis and tooling. Thus, \bosque is designed to encourage limited uses 
of recursion, increase the clarity of the recursive structure, and enable compilers/runtimes to avoid stack related issues~\cite{codecomplete}. 
To accomplish these goals \bosque borrows from the design of the \emph{async}/\emph{await} syntax and semantics~\cite{asyncawait} which is 
used to add structured asynchronous execution to a language. In this design the \emph{async}/\emph{await} keywords are used to explicitly identify 
functions that are asynchronous and when these functions are invoked. 

The \bosque language takes a similar approach by introducing the \cf{recursive} keyword which is used at both declaration sites to indicate a function/method 
is recursive and again at the call site so to affirm that the caller is aware of the recursive nature of the call. This feature is used in the binary-tree example 
when implementing the \cf{has} method. This method is declared as \cf{recursive} and later in the body at the callsite \cf{tchild.has[recursive](x)} the 
call is explicitly annotated as being potentially recursive. 

In \bosque the type-checker will process the call-graph for cycles and flag all caller-callee relations inside the \emph{same} cycle as requiring both annotations 
at the declaration and call-site. Thus, mutually recursive calls will require annotations on declarations and, recursive, call-sites as well.
These annotations primarily serve to make the, otherwise, implicitly recursive nature of these calls explicit in the code syntax. 
This provides clarity to the developer on which calls may involve recursion so they are not caught off-guard by the re-entrant nature of the code. This information 
also provides the compiler with the opportunity to convert an stack based call into a CPS form to avoid possible stack-overflows or enable static stack size computation 
for \emph{small-stacks}. The combination of explicit demarcation of recursive execution along with the ability to place strong pre/post conditions on these calls serve as limits on 
the complexity that recursion can introduce when reasoning about a block of code while still allowing recursion as an option for when functors cannot (or cannot reasonably) 
be used to express a computation.

\section{Simplicity and Clarity}
\label{sec:simplicity}

Given the design of the core IR (\autoref{sec:bsqir}) and the surface language (\autoref{sec:bsqsrc}) this section looks at how they resolve the challenges 
outlined in \autoref{sec:complexity} for each of the agents.

\subsection{Areas of Simplification}

\noindent
\paragraph{Immutable State and Local Reasoning:} The use of immutable value semantics and restrictions on exposing memory/object identity via 
equality operations ensures that the \bosque language is referentially transparent. As a result reasoning about the effects of a statement in general, 
and function/method calls in particular, can be done independently of the external context and using purely monotone reasoning. Specifically, 
no property that holds before some operation can be invalidated by the effect of the operation and the only parts of the program state that influence 
the operation are the argument values. 

\noindent
\paragraph{Explicit Behavior:} The lifting of implicit information to a textually explicit form with \emph{typedecls}, \emph{explicit flow typing}, 
and restricted lambda syntax ensure that the intent of blocks of code can be largely understood 
from their syntax. The addition of explicit support for \cf{pre}/\cf{post}, \cf{assert}, and \cf{invariant} declaration syntax lets us lift, otherwise 
diffuse and implicit information, into an explicit form that can be easily discovered. These features ensure that code can be reasoned about, 
primarily, by looking at the text and explicit declarations without the need to do extensive simulation of behavior, like a type-checking algorithm, 
a mutability check, or searching a codebase for diffuse information \eg every location a given type is constructed in the application. 

\noindent
\paragraph{Declarative Collection Processing and Recursion:} The elimination of loops in favor of \emph{collection functors} eliminates a 
major source of difficulty for symbolic analysis tooling. They also enhance the readability of the code by providing explicit and declarative ways of 
expressing operations on collections. The addition of explicit \cf{recursive} annotations provides a simple way to identify recursive calls to 
avoid \emph{unexpected reentrancy} issues and as a way to explicitly identify these calls for specialized processing when needed.

\noindent
\paragraph{Fully Deterministic Behavior:} Exhaustively specifying the semantics of each operation, including canonical orderings 
for associative containers, sort stability, and evaluation orders gives \bosque code a powerful property. Specifically, for any input there is a 
single, unique, and deterministic value that is the result. Thus, although \bosque programs are defined in terms of evaluation order and flow 
their semantics is isomorphic to the direct encoding in first-order logic (\autoref{sec:smallmv}).

\noindent
\paragraph{Atomic Data Operations:} The use of \emph{atomic constructors} and \emph{bulk-data operations} in \bosque, along with 
automatically checked invariants, makes it possible to ensure no value is ever in a partially defined or invalid state. This prevents 
accidental corruption and the construction of cyclic reference loops. Combined with the \cf{validate} support, these features ensure that 
when reasoning about code semantics, we can make strong guarantees about the properties that must hold at all program points. 

\noindent
\paragraph{Value Equality and Explicit Identity:} The restriction of equality comparisons to, primitive based, \emph{KeyType} values prevents 
the exposure of reference equality information (which would violate referential transparency). It also ensures that the definition of equality is 
uniform across an application and avoids the need to check for possible differences between, say, equality used in an associative container and 
equality as implemented in a \cf{==} operator. The elimination of semantically visible aliasing has additional benefits for runtime and compiler implementations.

\subsection{Reasoning Agents and Benefits}

\noindent
\paragraph{Human Developers:} The \bosque language provides a unique combination of features that eliminate various bug classes and 
simplify reasoning scenarios that humans find challenging. The primary area of improvement comes from regularizing application behavior 
in a way that reduces (or eliminates) special cases a human developer needs to keep in mind. For example there is no need to remember 
that sort order may change sometimes, that negation may overflow in one specific case (INT\_MIN), wonder if a call modifies global state, 
figure out what definition of equality will be used for a comparison. The second major benefit that \bosque provides for a human when reasoning about code 
is a strong bias for explicit intent expression. This includes the ability to explicitly specify logical invariants, the use of flow-type information, 
and the use of collection functors instead of depending on idiomatic loop structure to convey intent. These features enable 
developers to understand code explicitly instead of relying on (failable) intuition and patterns.

\noindent
\paragraph{Symbolic Analysis Tooling:} The \bosqueir representation is well-suited to supporting symbolic analysis tools. By construction, 
it eliminates major sources of complexity, including aliasing, mutability, and nondeterminism, and greatly simplifies other sources like 
inductive invariants. As a result it is, almost trivially, mappable to an efficiently solvable decidable/semi-decidable fragment of first 
order logic (\autoref{sec:smallmv}). Other symbolic analysis techniques, including abstract-interpretation based, also benefit from the 
reduced needs to perform strong-updates or frequently apply generic widening. Thus, these models are able to avoid getting 
''lost in the details'' of possible effects of an operation or losing information by making conservative assumptions in general cases. 
In practice this leads to increased scalability and precision of the analysis and, as a result, much more practical value from the 
tools/optimizations that they power.

\noindent
\paragraph{AI Agents:} As with human developers, the features in \bosque that explicitly encode intent in the syntax provides a major 
boost to LLM based agents. The features in \bosque also provide a richer set of information modalities for the 
models to use and extract information from. As seen in the \autoref{sec:aiprog} case study an agent working with \bosque code can use the 
textual language, evaluation of concrete values, and queries to symbolic tools that understand the declarative nature of invariants and 
assertions. These features improve the ability of the agent to extract useful information from the (limited) context it is given,
provide symbolic guardrails to limit the possibility of producing catastrophically wrong results, and allows the system to catch these 
mistakes quickly and minimize the impact when the agent does generate erroneous outputs.

\section{Case Studies}
\label{sec:casestudies}

In this section we examine how the features of \bosque impact mechanized development. This section uses two case studies, \emph{small model validation} and 
\emph{AI assisted programming} as representative studies to illustrate the potential for \bosque to power the future of software development.

\subsection{Implementation}
The \bosque language, including a compiler/type-checker, runtime, checker, synthesizer framework, and Cloud API specification framework, have all been implemented 
as open-source software and are publicly available\footnote{\bosque source code is available at \srclocation}. The initial implementation uses $30$kloc of TypeScript and $5$kloc of \bosque code. We expect this count 
to grow rapidly as the language moves from a collection of proof-of-concept components to a full-featured platform. There is active collaboration with colleagues 
at Microsoft to apply \bosque to technical challenges in API/Data Specification and software quality assurance.

\subsection{Small-Model Validation}
\label{sec:smallmv}
\paragraph{\bf Motivation}
Developers care deeply about the quality and reliability of the software they ship. However, there is a constant tension between time spent on 
quality and time spent building new features or addressing other client needs. For the majority of applications this calculation makes full-program verification 
an impractical option and, even with the needed resources, maintaining full-behavioral specifications is a Sisyphean task for most teams as they experience 
continuously changing business requirements and evolving feature sets. 

As a result (most) development teams are not interested in a system that performs full-proofs of correctness. Instead the sweet-spot is simple logical 
checks (asserts, pre/post, and data invariants) that can be written in the same language, and inline, as the application. Full proofs that these checks 
are always satisfied, are of course nice but developers often do not have time and technical ability to debug/resolve proof
failures, so more practically useful is generating inputs that trigger them if they can fail. In general the preference 
is for small inputs, or small reproductions that, are easy to debug and, are considered to exist for most possible failures (the small-model 
hypothesis~\cite{smallscope}). Under these constraints we want to create a checker that: 
\begin{enumerate}
\item Can be applied to any runtime \emph{or} user defined assert/invariant failure
\item Does not require any specialized annotations or developer knowledge of proof systems
\item Provides \emph{actionable} results in the form of a witness input when a failing condition is found
\end{enumerate}

This problem has been studied as a semi-decision procedure for $20+$ years in the form of Model Checking~\cite{CKY03}, Dynamic-Symbolic 
Analysis~\cite{dart}, concrete Fuzzing~\cite{afl}, and recently by formulating new (underapproximating) logics for modeling program 
semantics~\cite{incorrectness}. Despite the importance of the problem and the substantial amount of work on the topic it remains an unresolved 
challenge in practice.\\

\paragraph{Direct Solution with \bosque}
In contrast to other widely used languages\footnote{A notable exception is Elm~\cite{elm, highassurance}.} where the semantics are 
not efficiently encodable in first-order logic (FOL), due to features like loops, mutability, non-deterministic behaviors, etc. as 
identified in \autoref{sec:complexity}, \bosque can be converted in a direct manner into efficiently decidable FOL theories. 
The design restrictions on the \bosqueir core language enable us to map it, almost entirely, to efficiently 
decidable theories supported by a SAT-Module-Theory (SMT) solver~\cite{z3}. Operations on numbers, data-types, and functions all map to core 
decidable theories -- Integers, Constructors, Uninterpreted Functions, and Interpreted Functions. Strings and Sequences are used to model strings 
and collections. In Z3 the theories of Strings and Sequences are semi-decision procedures in the unbounded case. However, as we are interested in 
small-model inputs these are always bounded and become fully (and efficiently) decidable. As a result we can guarantee that an actionable 
(small) reproduction of a failure can be found if it exists and, from a practical perspective, this can be done automatically and efficiently in practice.

\paragraph{Example and Case Study}
To illustrate how this system works we show a (simplified) piece of code from a sample application, consisting of $2$Kloc of code, published by Morgan Stanley~\cite{morphirrepo}. 
The relevant type definitions definitions, \cf{SaleOrder} and \cf{SaleInfo} are in \autoref{fig:saleinfo}. The function in \autoref{fig:processing} takes a sale order, checks if there is available inventory to satisfy it, and then either 
accepts the order (adding it to the history) or returns \texttt{none} to indicate it was rejected. In this function there is one user defined property that 
needs to be checked, specifically that whenever an order is accepted the inventory must be reduced. This is clearly not a full, or even very complete specification, 
but in practice these types of sanity check conditions are very popular as they are effective in finding bugs and easy for developers to understand.

\begin{figure}
\begin{lstlisting}[language=Bosque]
function process(
    sales: SaleInfo, order: SaleOrder
): SaleInfo? 
    ensures $return != none ==> 
        $return@<SaleInfo>.available <= sales.available;
{
    if(sales.available < order.quantity) {
        return none;
    }
    else {
        return sales.{
            available=$available - order.quantity, 
            orders=$orders.pushBack(order)
        };
    }
}
\end{lstlisting}
\vspace{-4mm}
\caption{Order Processing from Trading App}
\label{fig:processing}
\end{figure}

Using the tooling that \bosque provides we can run the static checker over the application. This checker will enumerate every possible error in the application and then 
translate the relevant code to a (small-model) decidable fragment of logic. Each of these logical formula are passed to the Z3 SMT solver for either a satisfying assignment, 
which would be the failing input, or \emph{unsat} which indicates that there does not exist any small-model input that can trigger the error! When running 
the checker tool on the sample Fintech application we are able to produce a result for every error in under $0.2s$ per error (including process startup and loading 
SMTLIB files). For the \texttt{ensures} clause the tool reports that an error is possible and that is corresponds to the case where the \texttt{order} entity is:
\begin{lstlisting}[language=Bosque]
{
    id: "order_1",
    quantity: -1
}
\end{lstlisting}

If this concrete input is given to the application the ensures assertion will trigger as the negative \texttt{quantity} results in an increase in the \texttt{availability}. 
This is an error in the business logic, as \texttt{SaleOrder} is expected to always have a positive \texttt{quantity}. A developer can fix this bug by adding an invariant 
to the \texttt{SaleOrder} or changing the type of the \texttt{quantity} field to be a \texttt{BigNat} (instead of a \texttt{BigInt}). After either of the changes re-running the checker will report 
that there is no small input that can trigger this ensures clause.

This case study shows how the design of the \bosqueir representation enables the direct solution of a foundational software-engineering problem. The design 
of the intermediate language enables us to directly map code to efficiently decidable logics and avoid the complexities that have prevented the widespread 
use of these types of checkers in the past. Conversion of the full language semantics and checking of arbitrary user properties is among the most challenging 
reasoning problems in the SE tooling space. Thus, this is a clear demonstration that \bosque creates opportunities for advancement in the practical development of other 
tools, with simpler reasoning needs such as those based on abstract interpretation or dataflow analysis, as well.

\subsection{AI Assisted Programming}
\label{sec:aiprog}
AI assisted programming is in its early stages but several key challenges are already clear. The first is a need to provide guardrails for the code that 
these probabilistic agents generate. This is tied with the desire to provide multi-modal inputs for them to work with -- the program synthesis 
community~\cite{nlyze,flashfill,uist} has long looked at combinations of natural-language, formal specs, examples, and context as specifications 
for generating code. Finally, the current large-language-model (LLM) agents are most effective when dealing with text and, the more relevant 
information that can be hoisted into this representation, the more effective the agents are.

As an example consider the code below where we use a LLM agent to generate code for a function body that a developer has 
sketched out in TypeScript.
\begin{lstlisting}[language=Bosque]
/*Find the largest pair of values from the lists.*/
function maxPair(x: number[], y: number[]): 
    [number, number] 
{
    //generate the implementation code here 
} 
\end{lstlisting}

In a language such as TypeScript (or Java) the LLM must resolve the users intent solely from the the natural language in comments and (partial) code 
context. In our example passing this to GPT-4 (or Github Copilot) and asking for code completions produces multiple possible solutions the highest ranked 
in both cases is the following:
\begin{lstlisting}[language=Bosque]
let maxPair: [number, number] = [0, 0];
for (let i = 0; i < x.length; i++) {
    if (x[i] > maxPair[0]) {
        maxPair[0] = x[i];
        maxPair[1] = y[i];
    }
}
\end{lstlisting}
Interestingly this solution only looks for the maximum value in \texttt{x} and is unlikely to be the desired functionality. If we sample more solutions we 
also find versions that take the max from each list independently which is more likely to be the desired response but just from the source text it is 
difficult to make this choice with confidence.

In contrast \bosque has multiple features which are designed to make intent and specifications explicit in the source code. The expressive type system, including 
unions, nullable-types, and typedecls, along with the explicit syntactic support for pre/post conditions and invariants trivially exposes rich contextual information directly 
to the LLM. The code below shows the same signature but augmented with partial logical postconditions and examples of inputs and the corresponding outputs.
\begin{lstlisting}[language=Bosque]
/*Find the largest pair of values from the lists.*/
function maxPair(x: List<Int>, y: List<Int>): [Int, Int] 
    ensures x.contains($return.0);
    ensures y.contains($return.1);
    examples [
        [List{3, 2}, List{3, 5}] => [2, 5]
    ];
{
    defer; 
} 
\end{lstlisting}

With this extra contextual information the code completions generated are much higher quality. The top ranked solutions we extract include the 
following two candidates:
\begin{lstlisting}[language=Bosque]
(1) return [x.max(), y.max()];
(2) return List::zip<Int, Int>(x, y)
               .maxArg<Int>(fn(v) => v.0 + v.1);
\end{lstlisting}

Just having the ensures and examples as textual hints resulted in a substantial improvement in the generated code. The highest ranked output (\#1) 
is quite plausible. However, we can use the ensures and examples to further check the generated code. By running the examples as test cases 
we see that output (1) is not the desired result. Instead a slightly lower ranked output \#2\footnote{The exact rank of this version varies per run.} satisfies 
the example and the ensures clauses. Thus, after re-ranking we suggest output \#2, which in this case, is the actual desired output. A preliminary 
evaluation with manually blanking out bodies shows that the additional information available in \bosque (plus the absence of loops which are known 
problems for synthesis~\cite{tddsynth}) consistently improves results over simple text/code. 

This case study shows how \bosque enables the combination of natural language via comments and declarations, declarative constraints via the
\cf{ensures} clauses, and examples, in a form that a LLM can consume and use to drive the code generation task. These features 
provide a way to screen for invalid generations, by simply running the provided samples, and provide guardrails by validating the generated 
code against pre/post conditions and invariants (using methods like the previously described small-model validator). In addition to supporting the 
direct code synthesis task, this multi-modal interaction capability also opens up a variety of options for exploring user experiences and multi-round interactions 
as part of the code generation process~\cite{uist}.

\section{Related Work}
\label{sec:related}

Throughout this paper we have discussed the conceptual frameworks~\cite{structuredprogramming,silverbullet,adts,tecton} 
and language constructs~\cite{STL,loopmining,comega,typescript,durablefuncs,sml} that have motivated the 
development and the design of the \bosque language. Thus, this section focuses on topics related to the complexity 
issues identified and connections to other lines of research.

\noindent
\paragraph{\bf Invariant generation:} The problem of generating loop invariants goes back to the introduction of loops as a concept~\cite{hinvariants,finvariants}. 
Despite substantial work on the topic~\cite{invnonlinear,nipsinv,invdemand,invgen} the problem of generating precise loop invariants remains an open problem. This has severely limited 
the usability and adoption of formal methods in industrial development workflows. Notable successes include seL4~\cite{sel4}, CompCert~\cite{compcert}, and Everest~\cite{everest}. 
However, all of these systems required expertise in formal methods that is beyond what is available to most development teams. The \bosque language seeks to sidestep 
this challenge entirely by avoiding the presence of unconstrained iteration.

\paragraph{\bf Equality and Reference Identity:} Equality is a complicated concept in programming~\cite{lefthandequals}. Despite this complexity it has been under-explored 
in the research literature and is often defined based on historical precedent and convenience. This can result in multiple flavors of equality living in 
a language that may (or may not) vary in behavior and results in a range of subtle bugs~\cite{findbugs} that surface in surprising ways. 

Reference identity, and the equality relation it induces, is a particularly interesting example. Identity is often the desired version of equality for classic object-oriented
programming~\cite{lefthandequals} and having it as a default is quite convenient. However, in many cases a programmer desires equality based on values, or a 
primary key, or an equivalence relation and a default equality based on identity is, instead, a source of bugs. Further, the fact that it is based on memory addresses 
is a complication to pass-by-value optimizations of attempts to compile to non \emph{Von Neumann} architectures like FPGAs~\cite{fpga}.

\paragraph{\bf Alias Analysis:} The introduction of identity as an observable feature in a language semantics immediately pulls in the 
concept of aliasing. This is another problem that has been studied extensively over the years~\cite{palinear,pasolved,ptssa,paobjsens,padsa,paparallel} and remains an open 
and challenging problem. A major motivation for this work is, in a sense, to undo the introduction of reference identity and 
identify code where reference equality does not need to be preserved. This is critical to many compiler optimizations including 
classic transformations like scalar field replacement, conversion to pass-by-value, and copy-propagation~\cite{kennedyallen,muchnick}. 
This information is also critical to compiling to accelerator architectures like SIMD hardware~\cite{mesimd}.

\paragraph{\bf Frames and Ownership:}
The problem of aliasing is further compounded with the introduction of mutation. Once this is in the language the problem of computing frames~\cite{seplogic} 
and purity~\cite{purity} becomes critical. Often developers work around the problem of explicit frame reasoning by using an \emph{ownership}~\cite{owner1,owner2} 
discipline in their code. This may be a completely convention driven discipline or, more recently, may be augmented by runtime support such as smart pointers~\cite{STL} 
and type system support~\cite{rust,lineartypes,affinetypes,immutabletypes}.

\paragraph{\bf Concurrency, and Environmental Interaction:}
Reasoning in concurrent (parallel) applications with mutablility is a challenging problem. As all \bosque values are immutable the problem of Read-Write or Write-Write dependencies 
do not exist, so parallelism for performance can be done aggressively without concern for changing application behavior. Concurrency and non-determinism 
that result from environmental interaction such as user interactions, network, or external interaction with other processes are currently beyond the scope of 
\bosque. Instead it operates as a pure computation language that can be embedded by a host (or other language like Node.js modules~\cite{napi}) that manage 
async behavior and IO. A promising direction is integrating a core computation language (like \bosque) with an interaction focused language such as P/P\#~\cite{plang} 
that has sophisticated methods for analyzing and testing concurrency and environmental interactions~\cite{plangtest}.

\paragraph{\bf Incorrectness and Under Approximate Analysis:}
Incorrectness Logic~\cite{incorrectness} and other under approximate approaches~\cite{racerd}
represent an interesting and recent development in the design space of program 
analysis. These systems look to fuse the power of symbolic representations to capture many concrete states while under (rather than over) 
approximating reachability. Interestingly, one of the motivations for introducing Incorrectness Logic is that (p. 4) ``...the exact reasoning of 
the middle line of the diagram [strongest post semantics] is 
definable mathematically but not computable (unless highly incomputable formulae are used to describe the post).'' However, as shown in this 
paper, this middle line of exact and decidable semantics is practical to compute when the language semantics are designed 
appropriately. 

\paragraph{\bf Synthesis:} Program synthesis is an active topic of research but the need to reason about loops has limited the application of 
synthesis to mostly straight-line code. Work on code with loops has been more limited due to the challenge of reasoning about loops in code~\cite{mesimd,loopslater} 
and the difficultly synthesizers have constructing reasonable code that includes raw loop and conditional control-flow~\cite{tddsynth}. Thus, a language like \bosque, 
that provides high-level functors as primitives and can be effectively reasoned about opens new possibilities for program synthesis.

\section{Onward!}
This paper argues for a foundational re-conceptualization of the role of programming languages in the process of building software systems. Instead 
of being a set of increasingly powerful features and logical abstractions that a developer uses to formalize what is typed into a file, we advocate for them 
to be built as a substrate that is optimized for mechanization and reasoning tasks. This mindset led to revisiting many common assumptions about the features 
in a language and a drastic push for simplification of their semantics. \autoref{sec:complexity} enumerated these features, and the challenges they create 
for human/symbolic/statistical agents, while \autoref{sec:bsqir} and \autoref{sec:bsqsrc} show how a practical language can be designed to address 
these challenges.

To validate the effectiveness of the design in actually addressing the challenges identified we looked at two case studies that exercise different aspects of 
reasoning about an application. Both applications in \autoref{sec:casestudies} provide capabilities that are beyond the current state of the 
art in any mainstream programming ecosystem and, both, were built using variations on standard approaches. The key to enabling them was the 
ability to effectively perform reasoning on the application semantics!

With this initial success it is time to move Onward! Based on our experience with the language, case studies and proof-of-concept systems, we believe the 
core of the language is stable enough to build on. The compiler for \bosque is now being written in \bosque, collaborators 
are working with us on applying \bosque to solve critical technical challenges, and the potential for innovative tooling and platform research 
is massive\footnote{This project is fully open-source at: \srclocation}. We believe this is a unique 
opportunity for the academic and industrial communities to advance into a new era of programming languages that fully embraces the forces of mechanization, 
integration, AI driven coding, that are shaping the software development landscape. 

\if\techreport1
\section*{Acknowledgements}
This work is the result of many years of conversations, experiences, and thinking. I would like to give a special thanks to Ed Maurer, Gaurav Seth, Brian Terlson, 
Hitesh Kanwathirtha, Mike Kaufman, Todd Mytkowicz, and Earl Barr for all their thoughts and conversations. I would also like to thank Stephen Goldbaum and Richard Perris for their insights 
on how technology is impacting the FinTech sector. Finally, I want to acknowledge the the Node.js community for their innovation and willingness to experiment!
\fi

\bibliographystyle{plain}
\bibliography{bibfile} 

\end{document}